%% file: draft_Zc0.tex
\documentclass[twocolumn,showpacs,superscriptaddress]{revtex4-1}
\usepackage{graphicx}
\usepackage{units}
\usepackage{multirow}
\usepackage{bigstrut}
\usepackage{overpic}
\usepackage{array}
\usepackage{color}
\usepackage{subfigure}
\usepackage{amsmath}
\usepackage[displaymath, mathlines]{lineno}
\newcommand{\dst}{D^{*}}
\newcommand{\dstbar}{\bar{D}^{*}}

\newcommand{\dstzerobar}{\bar{D}{}^{*0}}
\newcommand{\dstzero}{D^{*0}}
\newcommand{\dstplus}{D^{*+}}
\newcommand{\dstminus}{D^{*-}}
\newcommand{\dplus}{D^{+}}

\newcommand{\dzero}{D^{0}}
\newcommand{\ee}{e^{+}e^{-}}

\newcommand{\gm}{\gamma}

\newcommand{\pip}{\pi^+}
\newcommand{\pizero}{\pi^0}
\newcommand{\pim}{\pi^-}

\newcommand{\mev}{\,\unit{MeV}}

\newcommand{\mevcc}{\,\unit{MeV}/c^2}
\newcommand{\gev}{\,\unit{GeV}}
\newcommand{\gevc}{\,\unit{GeV}/c}
\newcommand{\gevcc}{\,\unit{GeV}/c^2}

\newcommand {\ie}       {\emph{i}.\emph{e}.}

\newcommand {\Zn} {Z_{c}(4025)^{0}}
\newcommand {\Zp} {Z_{c}(4025)^{+}}

\begin{document}

\modulolinenumbers[2]

\title{\boldmath Observation of a neutral charmoniumlike state $Z_c(4025)^0$ in $e^{+} e^{-} \to (\dst \dstbar)^{0} \pizero$}

\input{authors_jun2015}

\begin{abstract}

We report a study of the process $e^{+} e^{-} \to (\dst \dstbar)^{0}
\pizero$ using $e^+e^-$ collision data  samples with integrated
luminosities of $1092\,\rm{pb}^{-1}$ at $\sqrt{s}=4.23\gev$ and
$826\,\rm{pb}^{-1}$ at $\sqrt{s}=4.26\gev$ collected with the BESIII
detector at the BEPCII storage ring. We observe a new neutral
structure near the $(\dst \dstbar)^{0}$ mass threshold in the
$\pizero$ recoil mass spectrum, which we denote as $\Zn$. Assuming a
Breit-Wigner line shape, its pole mass and pole width are determined to
be $(4025.5^{+2.0}_{-4.7}\pm3.1)\mevcc$ and $(23.0\pm 6.0\pm
1.0)\mev$, respectively. The Born cross sections of $e^{+}e^{-}\to
\Zn \pizero \to (\dst \dstbar)^{0}\pizero$ are measured to be
$(61.6\pm8.2\pm9.0)\,\rm{pb}$ at $\sqrt{s}=4.23\gev$ and
$(43.4\pm8.0\pm5.4)\, \rm{pb}$  at $\sqrt{s}=4.26\gev$. The first
uncertainties are statistical and the second are systematic.

\end{abstract}


\maketitle

Recent discoveries of new charmoniumlike states that do not fit
naturally with the predictions of the quark model have generated great
experimental and theoretical interests~\cite{snowmass2013}. Among these
so-called $``XYZ"$ particles are charged states with decay modes that
clearly demonstrate a structure consisting of at least four quarks,
including a $c \bar{c}$ pair. The first charged charmoniumlike state
$Z(4430)^{+}$ was discovered by Belle~\cite{Belle142001}. LHCb
confirmed the existence of this state. Belle determined its spin-parity
to be $1^{+}$~\cite{Belle074026}, which is supported by a new result
from LHCb\cite{LHCb222002}. Recently, the BESIII collaboration
observed four charged $Z_{c}$ states,
$Z_c(3885)^{\pm}$~\cite{BESIII022001},
$Z_c(3900)^{\pm}$~\cite{BESIII252001},
$Z_c(4020)^{\pm}$~\cite{BESIII242001}, and
$Z_c(4025)^{\pm}$~\cite{BESIII132001}, produced in $e^{+}e^{-} \to
\pi^{\mp} Z_{c}^{\pm}$. The observed decay channels are
$Z_c(3900)^{\pm}\to \pi^{\pm}J/\psi$, $Z_c(3885)^{\pm}\to
(D\bar{D}^{*})^{\pm}$, $Z_c(4020)^{\pm}\to \pi^{\pm} h_c$, and
$Z_c(4025)^{\pm}\to (D^{*}\bar{D}^{*})^{\pm}$. These states are close
to the $D\dstbar$ or $\dst\dstbar$ threshold. The  $Z_{c}(3900)^{\pm}$ was
also observed by Belle~\cite{Belle252002} and with \hbox{CLEO-c}
data~\cite{CLEO336}.

So far, the nature of these new states is still
elusive. Interpretations in terms of tetra-quarks, molecules,
hadro-charmonium, and cusp effects have been
proposed~\cite{Dianyong:036008,Z:054002,Z:194,Q:132003,F:054007,J:116004,Q:114009,X:1308,A:1310}. Searching
for their neutral partners in experiment is of great importance to
understand their properties, especially to identify their isospin
properties. Previously, based on CLEO-c data, evidence of a neutral
state $Z_c(3900)^0$ decaying to $\pizero J/\psi$~\cite{CLEO727} was
reported.  Recently, two neutral states, $Z_c(3900)^{0}$ and
$Z_{c}(4020)^{0}$, were discovered in their decays $Z_c(3900)^0 \to
\pizero J/\psi$ and $Z_c(4020)^0 \to \pizero h_c$ by
BESIII~\cite{BESIIIupdated,BESIII212002}. These can be interpreted as
the isospin partners of the $Z_c(3900)^{\pm}$ and
$Z_c(4020)^{\pm}$. Analogously, it is natural to search for the
neutral partner of the $Z_c(4025)^{\pm}$~\cite{BESIII132001} in its decay
to $(D^{*}\bar{D}^{*})^{0}$.

In this Letter, we report a search for the neutral partner of the
$Z_c(4025)^{\pm}$ through the reactions  $e^{+} e^{-} \to \dstzero
\dstzerobar (\dstplus\dstminus) \pizero$, as the charged
$Z_c(4025)^{\pm}$~\cite{BESIII132001} couples to $(\dst\dstbar)^{\pm}$
and has a mass close to the $(\dst\dstbar)^{\pm}$ mass threshold. We
denote the investigated final state products as
$(\dst\dstbar)^0\pizero$, where $\dst$ refers to $\dstzero$ or
$\dstplus$, and $\dstbar$ stands for their antiparticles. A partial
reconstruction method is applied to identify the $(\dst \dstbar)^0
\pizero$ final states. This method requires detection of a $D$ and a
$\bar{D}$  originating from $\dst$ and $\dstbar$ decays of $\dst \to D \pi$ and $D \gamma$, and the
$\pizero$ from the primary production (denoted as the \emph{bachelor}
$\pizero$). The data sample analyzed corresponds to $\ee$ collisions
with integrated luminosities of $1092\,\rm{pb}^{-1}$ at
$\sqrt{s}=4.23\gev$ and $826\,\rm{pb}^{-1}$ at
$\sqrt{s}=4.26\gev$~\cite{BESIII03408} collected with the BESIII
detector~\cite{BESIIIdetector} at the BEPCII storage
ring~\cite{PEPCII}.

BESIII is a cylindrically symmetric detector, which from inner to
outer parts consists of the following components: a Helium-gas based
multilayer drift chamber (MDC), a time-of-flight counter (TOF), a
CsI(Tl) crystal electromagnetic calorimeter (EMC), a 1-Tesla
superconducting solenoid magnet and a 9-layer RPC-based muon chamber
system. The momentum resolution for charged tracks in the MDC is
$0.5\%$ at a momentum of $1\gevc$. The energy resolution for photons
in EMC with energy of $1\gev$ is $2.5\%$ for the center region
(barrel) and $5\%$ for the rest of the detector (endcaps). For charged
particle identification (PID), probabilities $\mathcal{L}(h)$ for
particle hypotheses $h=\pi$ or $K$ are evaluated based on the normalized
energy loss $\mathrm{d}E/\mathrm{d}x$ in the MDC and the time of
flight in the TOF. More details on the BESIII spectrometer can be
found in Ref.~\cite{BESIIIdetector}.

To optimize data-selection criteria, understand backgrounds and
estimate the detection efficiency, we simulate the $e^{+}e^{-}$
annihilation processes with the {\sc kkmc} algorithm~\cite{KKMC},
which takes into account continuum processes, initial state radiation
(ISR) return to $\psi$ and $Y$ states, and inclusive $D_{(s)}$
production.  The known decay rates are taken from the Particle Data
Group (PDG)~\cite{pdg2014} and decays are modeled with {\sc
  evtgen}~\cite{Lange:2001uf}. The remaining decays are simulated with
the {\sc lundcharm} package~\cite{Chen:2000tv}.  The non-resonant, three-body phase
space (PHSP) processes $e^{+}e^{-} \to D^{*} \bar{D}^{*} \pizero$ are
simulated according to uniform distributions in momentum phase
space. We assume that $\Zn$ has spin-parity of $1^{+}$ by considering the measurements of other $Z$ resonances~\cite{Belle074026,LHCb222002} and the signal
process $e^{+}e^{-} \to \Zn \pizero$ followed by
$\Zn \to (D^{*} \bar{D}^{*})^0$ proceeds in $S$ waves. The $\dst$ is
required to decay inclusively according to its decay branching ratios
from PDG~\cite{pdg2014}. The $\dplus$ is required to decay into
$K^{-} \pip \pip$ while $\dzero$ is required to decay into
$K^{-} \pip$, $K^{-} \pip \pizero$ and $K^{-} \pip \pip \pim$. These
decay modes are the ones used to reconstruct $D$ mesons~\cite{implied}. All
simulated MC events are fed into a {\sc geant4}-based~\cite{GEANT4} software package,
taking into account detector geometry and response.

The charged tracks of $K^-$ and $\pi^\pm$ are reconstructed in the
MDC. For each charged track, the polar angle $\theta$ defined with
respect to the $e^+$ beam is required to satisfy
$|\rm{cos\theta}|<0.93$. The closest approach to the $e^{+} e^{-}$
interaction point is required to be within $\pm 10~\rm{cm}$ along the
beam direction and within $1~\rm{cm}$ in the plane perpendicular
to the beam direction. A track is identified to be a $K(\pi)$ when the
PID probabilities satisfy $\mathcal{L}(K)>\mathcal{L}(\pi)$
($\mathcal{L}(K)<\mathcal{L}(\pi)$), according to the information from
$\mathrm{d}E/\mathrm{d}x$ and TOF.

The $\pi^0$ candidates are reconstructed by combining pairs of photons
reconstructed in the EMC that are not associated with charged
tracks. For each photon, the energy deposition in the EMC barrel
region is required to be greater than $25\mev$, while in the end-cap
region, it must be greater than $50\mev$, due to the different detector resolution and probabilities of reconstructing a fake photon. To suppress electronics
noise and energy deposits unrelated to the event, the EMC cluster time is restricted to be
within a 700 ns window near the event start time. The invariant mass
of any pair of photons $M(\gm\gm)$ is required to within
$(0.120, 0.145)\gevcc$ and constrained to the nominal $\pizero$
mass. The kinematics of the two photons are updated according to the
constraint fit.

We consider all possible combinations of selected charged tracks and
$\pizero$ to form $D$ candidates. The charged tracks from a $D$ decay
candidate are required to originate from a common vertex. The $\chi^{2}_{\rm{VF}}$
of the vertex fit is required to satisfy $\chi^{2}_{\rm{VF}}<100$. We
constrain the reconstructed masses of the final state particles to the
corresponding $D$ nominal masses and require $\chi^{2}_{\rm{KF}}(D)$
for the kinematic fit to be less than $15$ for the final states of $D$
decays including charged tracks only, and less than $20$ for the final
state including $\pizero$. We select signal event candidates which
consist of at least one pair of $D\bar{D}$ candidates that do not
share particles in the final state. If there is more than one
pair of $D \bar{D}$ candidates in an event, only the one with the
minimum $\chi^{2}_{\rm{KF}}(D) +\chi^{2}_{\rm{KF}}(\bar{D}) $ is kept
for further analysis.

We reconstruct the bachelor $\pizero$ from the remaining photon
showers that are not assigned to the $D\bar{D}$ pair. To further
reject backgrounds, each photon candidate originating from the
bachelor $\pizero$ is required not to form a $\pizero$ candidate with
any other photon in the event. A mass constraint of the two photons to
the $\pizero$ nominal mass is implemented and the corresponding fit
quality is required to satisfy $\chi^{2}_{\rm{KF}}(\pizero)<20$.  To
reject background for the bachelor $\pizero$ from $\dst\to D\pizero$
decays, we require the $D \pizero$ invariant mass to be greater than
$2.02\gevcc$.

\begin{figure}
\centering
\begin{overpic}[width=0.49\linewidth]{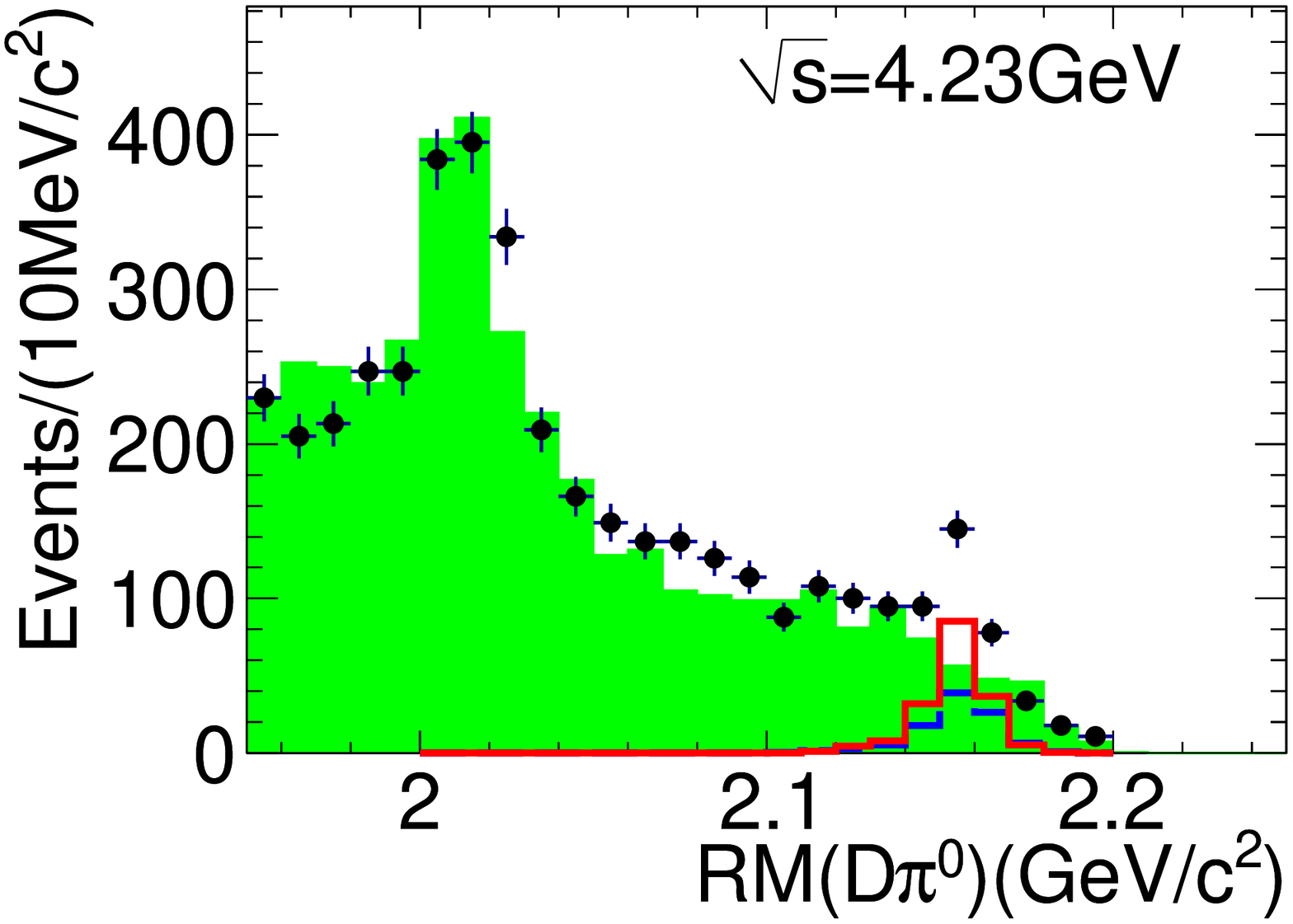}
\put(85,30){\textbf{(a)}}
\end{overpic}
\begin{overpic}[width=0.49\linewidth]{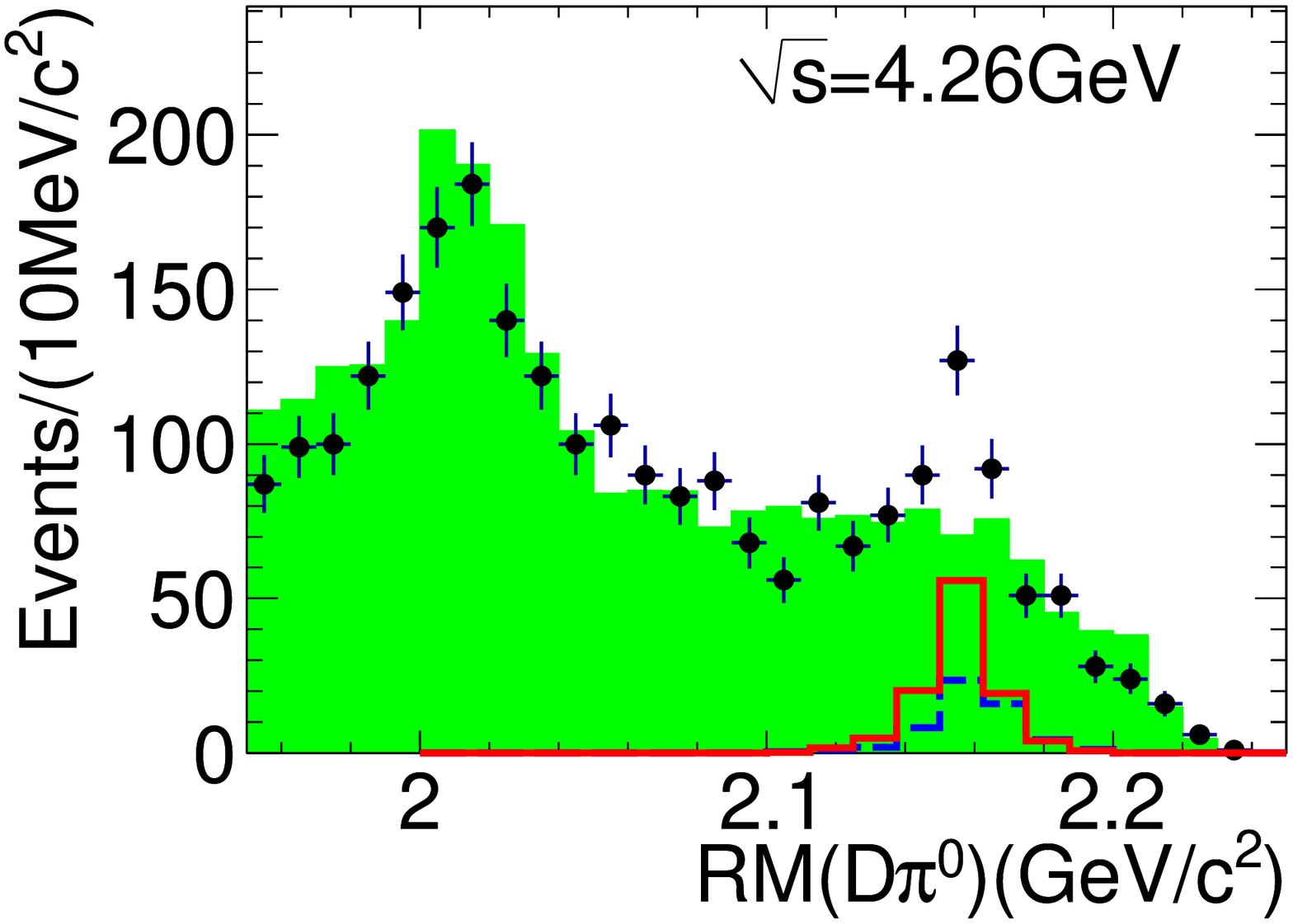}
\put(85,30){\textbf{(b)}}
\end{overpic}
\begin{overpic}[width=0.49\linewidth]{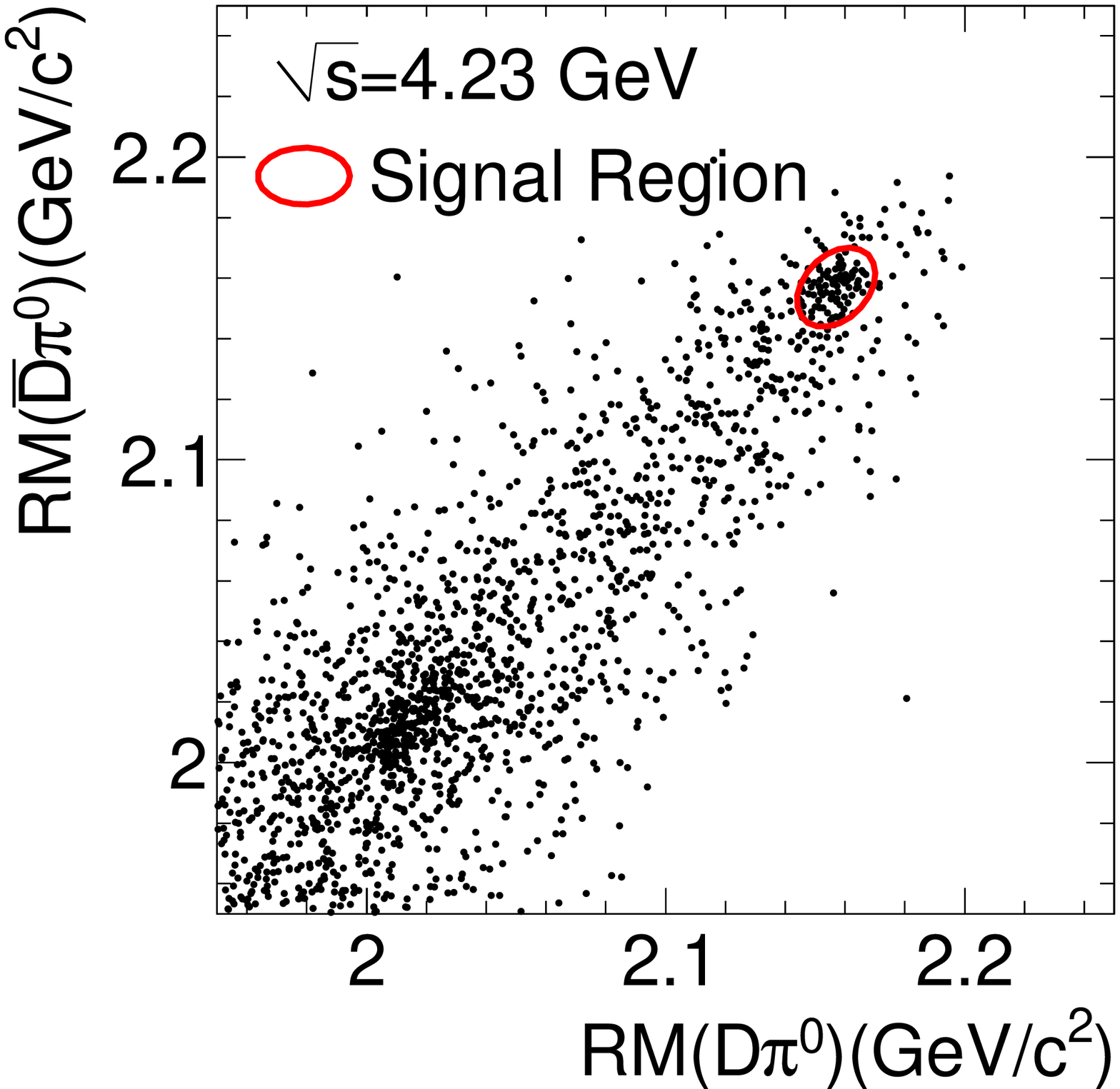}
\put(85,40){\textbf{(c)}}
\end{overpic}
\begin{overpic}[width=0.49\linewidth]{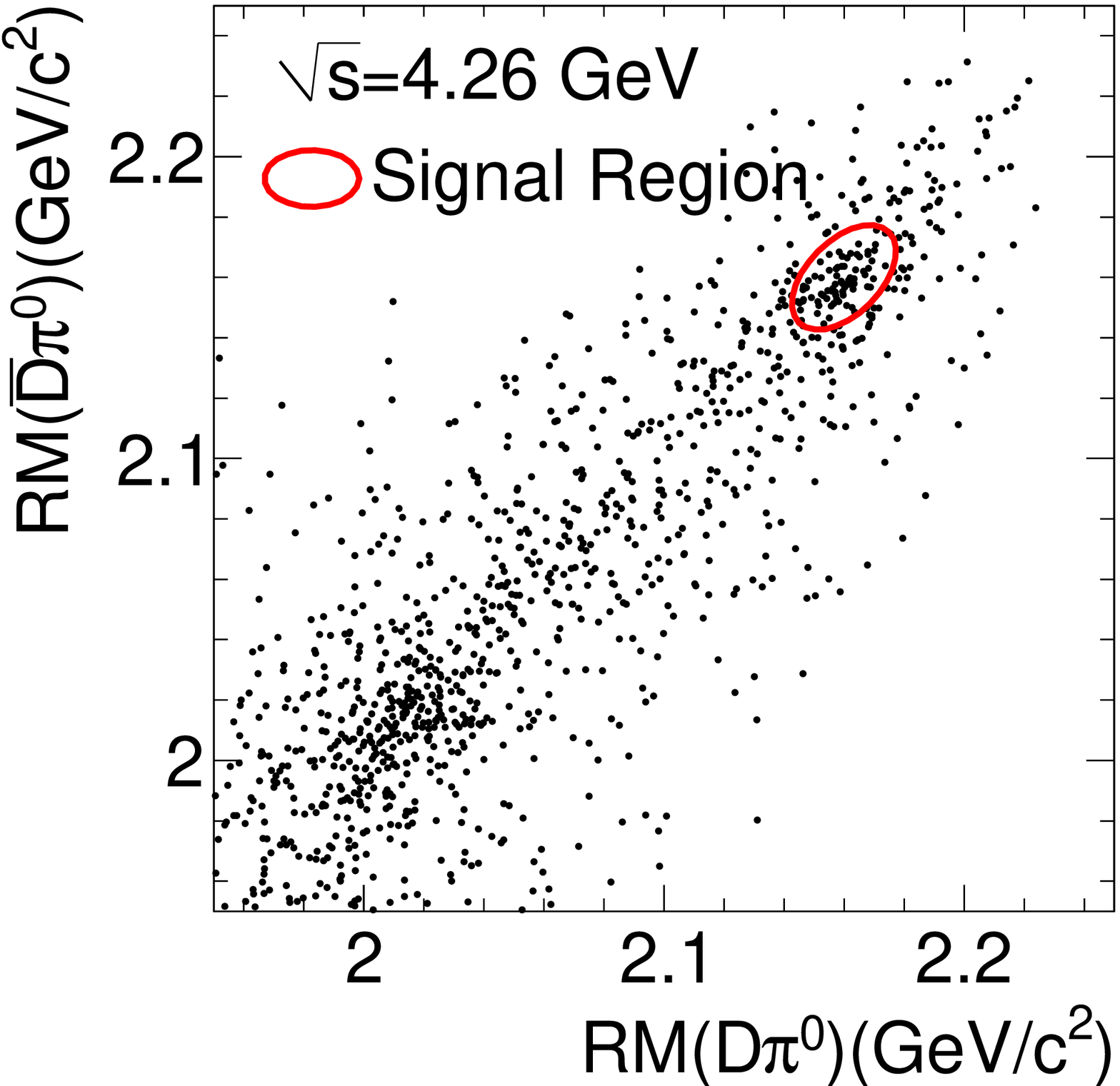}
\put(85,40){\textbf{(d)}}
\end{overpic}
\caption{Distributions of $RM(D \pi^{0})$ at $\sqrt{s}=4.23\gev$ (a) and
  $\sqrt{s}=4.26\gev$ (b). Points with error bars are data and the
  shaded histograms represent the inclusive backgrounds in MC
  simulations. The soild line and the dashed line are the $\Zn$
  signal shape and the PHSP shape with arbitrary normalization,
  respectively. The third row gives the scatter plot of $RM(D\pizero)$
  versus $RM(\bar{D}\pizero)$ at $\sqrt{s}=4.23\gev$ (c) and
  $\sqrt{s}=4.26\gev$ (d) , where the solid ovals indicate the signal
  regions.}
\label{fig:rmDpi0}
\end{figure}

To identify the decay products of the signal process
$e^{+}e^{-} \to \dst\dstbar\pizero$, we plot the recoil mass spectra
of $D \pizero$ ($RM(D \pizero)$), as shown in
Fig.~\ref{fig:rmDpi0}. The peaks around $2\gevcc$ correspond to the
process $e^{+}e^{-}\to D \dstbar \pizero$ with a missing
$\dstbar$. Besides these peaks, we see clear bumps around
2.15$\gevcc$ in data. These bumps are consistent with the MC simulations of
the $\dst \dstbar \pizero$ final state. The peak position roughly
corresponds to the sum of the mass of $\dst$ and the mass of a $\pi$,
since the $\pi$ originating from $\dst$ is soft and is not used in the
computation of the recoil mass. The backgrounds beneath the bumps are
mostly from ISR production of $\dst \dstbar$ process. Other processes,
such as $e^{+}e^{-} \to \dst \bar{D}^{**} \to \dst \dstbar \pizero$, are
expected to be absent according to simulation studies. This is understandable
because the process $D_{0}^{*}(2400) \to \dst \pizero$ is forbidden
due to the conservation of spin-parity. $D_{1}^{*}(2420)^{0}$
($D_{2}^{*}(2460)^{0}$) is narrow, and the sum of the mass of
$D_{1}^{*}(2420)^{0}$ ($D_{2}^{*}(2460)^{0}$) and $D^{*}$ is much
larger than $4.26\gev$. To extract the signals, we keep events within
the two-dimensional oval regions in the distributions of
$RM(D\pizero)$ and $RM(\bar{D}\pizero)$ shown in
Fig.~\ref{fig:rmDpi0}(c,d). We choose the specific dimensions due to different resolutions at different momentum phase spaces at two energy points. They are determined according to MC simulation.

\begin{figure}
\centering
\begin{overpic}[width=0.95\linewidth]{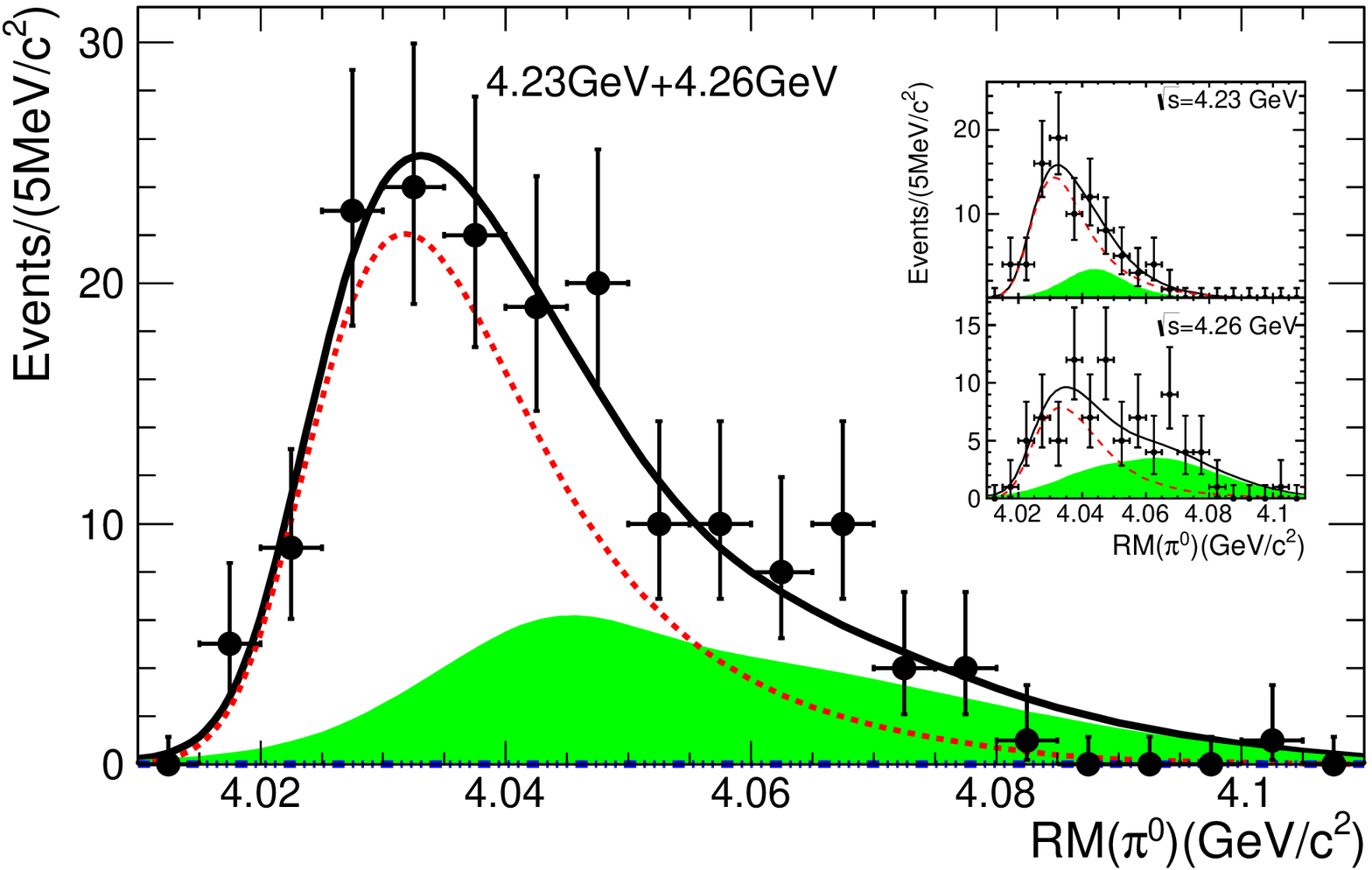}
\put(15,50){\textbf{(a)}}
\end{overpic}
\begin{overpic}[width=0.95\linewidth]{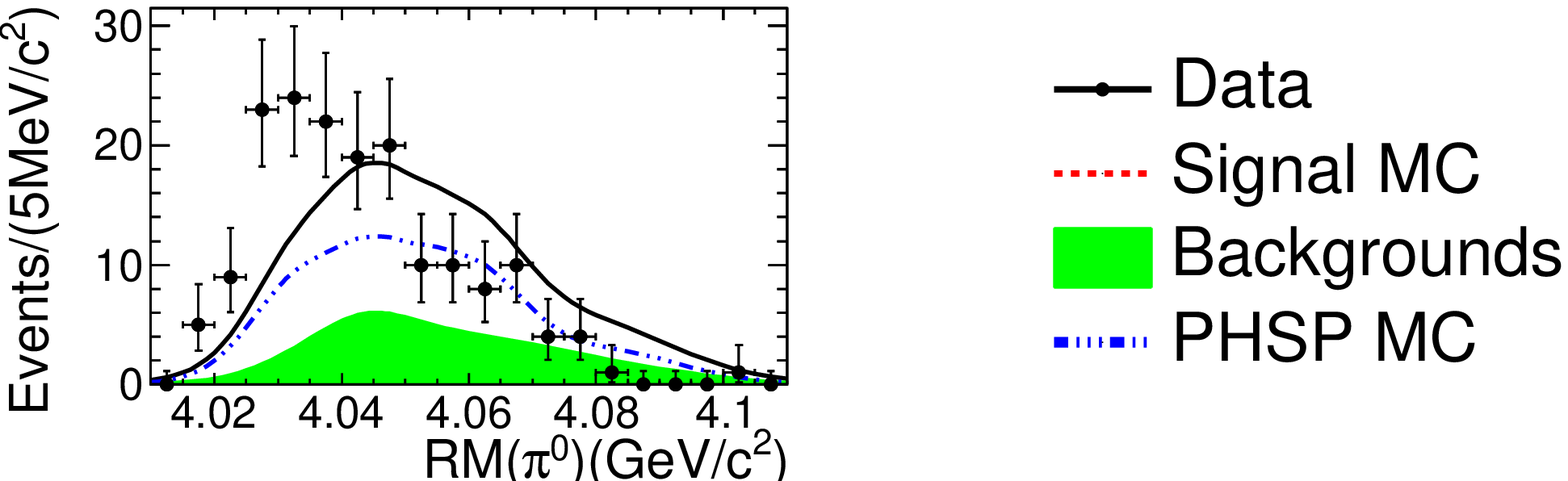}
\put(35,17){\textbf{(b)}}
\end{overpic}

\caption{Fits to $RM(\pizero)$. (a) A fit to background, PHSP
  and $\Zn$ signal process for the combination of all data (main
  plot), and the two collision energies separately (inset
  plots). (b) Fits using only the inclusive background and PHSP. Points
  with error bars are data, solid line is the sum of fit functions,
  dotted line stands for the $\Zn$ signals, filled area represents
  inclusive backgrounds, and dash-dotted line is the PHSP process.}
\label{fig:simultaneousfit}
\end{figure}

The selected events are used to produce the recoil mass distribution
of the bachelor $\pizero$ ($RM(\pizero)$), shown in
Fig.~\ref{fig:simultaneousfit}. We observe enhancements in the
$RM(\pizero)$ distribution over the inclusive backgrounds for both
data samples, which can not be explained by three-body non-resonant
processes. We assume the presence of an $S$-wave
Breit-Wigner resonance structure (denoted as $\Zn$) with a
mass-dependent width, using the form given in Ref.~\cite{LTP:114013}:

$$\left|\frac{1}{M^{2}-m^{2}-i \cdot m(\Gamma_{1}(M)+\Gamma_{2}(M))/c^{2}}\right|^{2}\cdot p_{k} \cdot q_{k},$$
$$ \mathrm{and} ~~ \Gamma_{k}(M)=f_k\cdot\Gamma\cdot \frac{p_{k}}{p_{k}^{*}} \cdot \frac{m}{M} ~~ (k=1,2).$$
Here, $k=1$ and 2 denote the neutral channel
$\Zn \to \dstzero \dstzerobar$ and the charged channel
$\Zn\to \dstplus \dstminus$, respectively. $f_k$ is the ratio of the
partial decay width for channel $k$. $M$ is the reconstructed mass,
$m$ is the resonance mass and $\Gamma$ is the resonance
width. $p_{k}(q_{k})$ is the $D^{*}$($\pizero$) momentum in the rest
frame of the $D^{*}\bar{D}^{*}$ system (the initial $e^{+}e^{-}$
system) and $p_{k}^{*}$ is the momentum of $D^{*}$ in the $\Zn$ rest
frame at $M=m$.  We assume that $\Zn$ decay rates to the neutral
channel and the charged channel are equal, \ie, $f_k=0.5$, based on
isospin symmetry.

We perform a simultaneous unbinned maximum likelihood fit to the spectra of
$RM(\pizero)$ at $\sqrt{s}=4.23$ and $4.26\gev$.  The
signal shapes are taken as convolutions of the efficiency-weighted
Breit-Wigner functions with resolution functions obtained from MC
simulations. The detector resolutions are $4\mev$ at
$\sqrt{s}=4.23\gev$ and $4.5\mev$ at $\sqrt{s}=4.26\gev$. Backgrounds
are modeled with kernel-estimated non-parametric
shapes~\cite{Cranmer:2000du} based on the inclusive MC, and their
magnitudes are fixed according to the simulations, since the inclusive
MC samples well describe the background. The shape of the PHSP process
is adopted from MC simulations. We combine the data at
$\sqrt{s}=4.23\gev$ and $\sqrt{s}=4.26\gev$ together, as shown in
Fig.~\ref{fig:simultaneousfit}. The fit determines $m$ and
$\Gamma$ to be $(4031.7 \pm 2.1) \mevcc$ and $(25.9 \pm 8.8)\mev$,
respectively. The corresponding pole position
$m_{\rm{pole}}(\Zn)-i\frac{\Gamma_{\rm{pole}}(\Zn)}{2}$ is calculated
to be
\begin{displaymath}
m_{\rm{pole}}(\Zn)=(4025.5^{+2.0}_{-4.7})\mevcc,
\end{displaymath}
\begin{displaymath}
\Gamma_{\rm{pole}}(\Zn)=(23.0\pm6.0)\mev.
\end{displaymath}
The significance with systematic errors is estimated by comparing the likelihoods of the fits
with and without the $\Zn$ signal component included. The likelihood
difference is $2 \Delta \ln L = 45.3$ and the difference of the number
of free parameters is $4$. When the systematic uncertainties are taken into account with the assumption of Gaussian distribution, the significance is estimated to be $5.9 \sigma$.

The Born cross section {\small
  $\sigma(e^{+}e^{-} \to \Zn \pizero \to
  (\dstzero\dstzerobar+\dstplus\dstminus)\pizero)$}
is calculated from the equation
\begin{displaymath}
\sigma = \frac{ n_{\rm sig}} { \mathcal{L} ( f_1\mathcal{B}_{1}\varepsilon_{1} + f_2\mathcal{B}_{2} \varepsilon_{2} ) (1+\delta) (1+\delta_\text{vac})},
\end{displaymath}
where $\mathcal{L}$ is the integrated luminosity, $\varepsilon_{1}$
($\varepsilon_{2}$) is the detection efficiency of the neutral
(charged) channel, $f_1$ ($f_2$) is the ratio of the cross section of
the neutral (charged) channel to the sum of the both channels,
$\mathcal{B}_{1}$ ($\mathcal{B}_{2}$) is the product branching
fraction of the neutral (charged) $\dst$ decays to the final states we
detected.  $(1+ \delta)$ is the radiative correction factor and
$(1+\delta_\text{vac})$ is the vacuum polarization factor. From the
simultaneous fit, we obtain $69.5\pm9.2$ signal events at
$\sqrt{s}=4.23\gev$ and $46.1\pm8.5$ signal events at
$\sqrt{s}=4.26\gev$.  $(1+ \delta)$ is calculated to be $0.744$ at
$\sqrt{s}=4.23\gev$ and $0.793$ at $\sqrt{s}=4.26\gev$ to the second
order in QED~\cite{Kuraev466}, where the input line shape of the cross
section is assumed to be the same as for
$e^{+}e^{-} \to (\dst \dstbar)^{+} \pim$, as extracted directly from
BESIII data. $(1+\delta_\text{vac})$ is given as 1.054 following the
formula in Ref.~\cite{arxi11074683}. The efficiency
$\varepsilon_{1}$ ($\varepsilon_{2}$) is determined to be $1.49 \%$
($3.87 \%$) at $\sqrt{s}=4.23\gev$ and $1.84 \%$ ($4.37 \%$) at
$\sqrt{s}=4.26\gev$.  Thus, the cross sections are measured to be
$ (61.6\pm8.2)\, \rm{pb}$ and $(43.4\pm8.0)\, \rm{pb}$ at
$\sqrt{s}=4.23$ and 4.26$\gev$, respectively. The contribution of the
PHSP process is found to be negligible according to the fit.

\begin{table}
  \begin{center}
    \caption{Summary of systematic uncertainties on the $\Zn$
      resonance parameters and cross sections $\sigma_{4230}$ at
      $\sqrt{s}=4.23\gev$ and $\sigma_{4260}$ at
      $4.26\gev$. ``$\cdots$'' means the uncertainty is negligible.
      The total systematic uncertainty is taken as the root of the
      quadratic sum of the individual uncertainties.}
  \begin{tabular}{l|ccccc}

      \hline \hline
      \footnotesize{Source}   & $m$\footnotesize{($\mevcc$)}  & $\Gamma$\footnotesize{($\mev$)}  &  $\sigma_{4230}$\footnotesize{(\%)} &  $\sigma_{4260}$\footnotesize{(\%)}  \\ \hline
      \footnotesize{Tracking}  &  & & 5 & 5 \\
      \footnotesize{Particle ID} &  &  & 5 & 5 \\
      \footnotesize{$\pi^{0}$ reconstruction}  & & & 4 & 4\\
      \footnotesize{Photon veto}  & & &4.2 & 4.2\\
      \footnotesize{Mass  scale}  &   2.6 &  &  &  \\
      \footnotesize{Detector resolution}  & 0.2  & 0.1 & 0.3  & 0.5   \\
      \footnotesize{Backgrounds}  & 0.6 & 0.2 & 5.6 & 5.4  \\
      \footnotesize{Oval cut} & 1.5 &$1.0$  & 4.2 & 2.0  \\
      \footnotesize{Fit range}  & $\cdots$ & 0.1 & 0.3 & 0.5  \\
      \footnotesize{$D^{*}\bar{D}^{*}\pizero$ line shape} &$\cdots$ & $\cdots$ & 6.0 & 3.0 \\
      \footnotesize{Luminosity} &  & & 1  &1 \\
      \footnotesize{$\mathcal{B}_{1}$ and $\mathcal{B}_{2}$} &$\cdots$  &$\cdots$ & $6.5$ & $5.3$   \\
      \footnotesize{Isospin violation} &$\cdots$ &0.2 & 0.3 & 0.2  \\
      \footnotesize{Vacuum polarization}  &   &   & 0.5 & 0.5  \\
          \hline
          \footnotesize{Total} & $3.1$ & $ 1.0$ & $ 14.6$  & $12.5$   \\
        \hline\hline
    \end{tabular}
    \label{tab:sys_err}
  \vspace{-0.6cm}
  \end{center}
\end{table}

Sources of systematic uncertainties in the measurement of the $\Zn$
resonance parameters and cross sections are listed in
Table~\ref{tab:sys_err}. Uncertainties of tracking and PID are each
$1\%$ per track~\cite{BESIII112005}. The uncertainty of the $\pizero$
reconstruction efficiency is $4\%$~\cite{BESIII052005}. We study the
photon veto by fitting the recoil mass of $D \pizero$ with and without
this veto in selecting the control sample of
$\ee\to(\dst\dstbar)^0\pizero$ in data. The efficiency-corrected
signal yields are used to extract the cross section, and the
corresponding change is taken into account as the systematic error
introduced by this requirement. The systematic uncertainties are
determined to be $4.2\%$ for both data samples. The mass-scale
uncertainty for the $\Zn$ mass is estimated with the mass shift
(comparison between the PDG nominal values and fit values) of
$RM(D\pizero)$ in the control sample
$e^{+}e^{-} \to D \bar{D} \pizero$ and of $RM(D)$ in the control
sample of $e^{+}e^{-} \to D \bar{D}$. To be conservative, the largest
difference of the two mass shifts, $2.6\mevcc$, is assigned as the
systematic uncertainty due to the mass scale. The systematic
uncertainty from backgrounds is estimated by leaving free the
magnitudes in the fit
and making different choices in non-parametric kernel-estimation of
the background events to account for the limited precision in MC simulation~\cite{kernel}. We change the oval cut criteria and take the
largest difference as the systematic uncertainty. Since the line shape
will affect the efficiency and $(1+\delta)$, to evaluate the
systematic uncertainties with respect to the input
$D^{*} \bar{D}^{*} \pizero$ line shape, we change its shape based on uncertainties of the observed $\dstplus \dstzerobar \pim$ cross
section. Branching fractions $\mathcal{B}_{1}$ and $\mathcal{B}_{2}$
are used in calculating the cross sections and the uncertainties of
the world average results are included as part of the systematic
uncertainty.

Other items in Table~\ref{tab:sys_err} have only minor effects on the
precision of the results. We change the fitting ranges in the
$RM(\pizero)$ spectrum and take the largest difference as the
systematic uncertainty. The uncertainties due to detector resolution
are accounted for by varying the widths of the smearing functions. The
uncertainty of integrated luminosity is determined to be $1\%$ by
measuring large angle Bhabha events~\cite{BESIII242001}. We vary the
ratio $f_k$ from 0.4 to 0.6 to take into account potential isospin
violation between the neutral and charged processes. The corresponding
changes are assigned as systematic uncertainties. The systematic
uncertainty of the vacuum polarization factor is
$0.5\%$~\cite{arxi11074683}.

In summary, using $\ee$ annihilation data at $\sqrt{s}=4.23$ and
$4.26\gev$, we observe enhancements in the $\pizero$ recoil mass
spectrum in the process
$e^{+}e^{-} \to \dstzero \dstzerobar (\dstplus \dstminus)
\pi^{0}$.
Assuming that the enhancement is due to a neutral charmoniumlike state
decaying to $\dst \dstbar$ and it has spin-parity of $1^{+}$, the mass and width of its pole position
are determined to be
$m_{\rm{pole}}(\Zn)= (4025.5^{+2.0}_{-4.7}\pm3.1)\mevcc$ and
$\Gamma_{\rm{pole}}(\Zn)= (23.0\pm6.0 \pm 1.0)\mev$, respectively. The
Born cross section {\small
  $\sigma(e^{+}e^{-} \to \Zn \pizero
  \to(\dstzero\dstzerobar+\dstplus\dstminus)\pizero)$}
is measured to be $(61.6\pm8.2 \pm9.0) \,\rm{pb}$ at
$\sqrt{s}=4.23\gev$ and $(43.4\pm8.0 \pm 5.4)\, \rm{pb}$ at
$\sqrt{s}=4.26\gev$.  Hence, we estimate the ratio
$\frac{\sigma(\ee \to \Zn
  \pizero\to(\dst\dstbar)^0\pizero)}{\sigma(\ee\to \Zp
  \pi^-\to(\dst\dstbar)^+\pi^-)}$
to be compatible with unity at $\sqrt{s}=4.26\gev$, which is expected
from isospin symmetry. In addition, the $\Zn$ has mass and width very
close to those of the $Z_c(4025)^{\pm}$, which couples to
$(\dst \dstbar)^{\pm}$~\cite{BESIII132001}. Therefore, the observed
$\Zn$ state in this Letter is a good candidate to be the isospin
partner of $Z_c(4025)^{\pm}$.


The BESIII collaboration thanks the staff of BEPCII and the IHEP
computing center for their strong support. This work is supported in
part by National Key Basic Research Program of China under Contract
No.~2015CB856700; National Natural Science Foundation of China (NSFC)
under Contracts Nos.~11125525, 11235011, 11275266, 11322544, 11335008, 11425524;
the Chinese Academy of Sciences (CAS) Large-Scale Scientific Facility
Program; the CAS Center for Excellence in Particle Physics (CCEPP);
the Collaborative Innovation Center for Particles and Interactions
(CICPI); Joint Large-Scale Scientific Facility Funds of the NSFC and
CAS under Contracts Nos.~11179007, U1232201, U1332201; CAS under
Contracts Nos.~KJCX2-YW-N29, KJCX2-YW-N45; 100 Talents Program of CAS;
INPAC and Shanghai Key Laboratory for Particle Physics and Cosmology;
German Research Foundation DFG under Contract No.~Collaborative
Research Center CRC-1044; Istituto Nazionale di Fisica Nucleare,
Italy; Ministry of Development of Turkey under Contract
No.~DPT2006K-120470; Russian Foundation for Basic Research under
Contract No.~14-07-91152; U.S.~Department of Energy under Contracts
Nos.~DE-FG02-04ER41291, DE-FG02-05ER41374, DE-FG02-94ER40823,
DESC0010118; U.S.~National Science Foundation; University of Groningen
(RuG) and the Helmholtzzentrum fuer Schwerionenforschung GmbH (GSI),
Darmstadt; WCU Program of National Research Foundation of Korea under
Contract No.~R32-2008-000-10155-0.


\end{document}

%% file: authors_jun2015.tex
\author{
  \begin{small}
    \begin{center}
      M.~Ablikim$^{1}$, M.~N.~Achasov$^{9,f}$, X.~C.~Ai$^{1}$,
      O.~Albayrak$^{5}$, M.~Albrecht$^{4}$, D.~J.~Ambrose$^{44}$,
      A.~Amoroso$^{48A,48C}$, F.~F.~An$^{1}$, Q.~An$^{45,a}$,
      J.~Z.~Bai$^{1}$
% ...
      \\
      \vspace{0.2cm}
      (BESIII Collaboration)\\
      \vspace{0.2cm} {\it
        $^{1}$ Institute of High Energy Physics, Beijing 100049, People's Republic of China\\
% ...
$^{4}$ Bochum Ruhr-University, D-44780 Bochum, Germany\\
        $^{5}$ Carnegie Mellon University, Pittsburgh, Pennsylvania 15213, USA\\
% ...
$^{9}$ G.I. Budker Institute of Nuclear Physics SB RAS (BINP), Novosibirsk 630090, Russia\\
% ...
$^{44}$ University of Rochester, Rochester, New York 14627, USA\\
        $^{45}$ University of Science and Technology of China, Hefei 230026, People's Republic of China\\
% ...
$^{48}$ (A)University of Turin, I-10125, Turin, Italy; (B)University of Eastern Piedmont, I-15121, Alessandria, Italy; (C)INFN, I-10125, Turin, Italy\\
% ...
$^{a}$ Also at State Key Laboratory of Particle Detection and Electronics, Beijing 100049, Hefei 230026, People's Republic of China\\
% ...
$^{f}$ Also at the Novosibirsk State University, Novosibirsk, 630090, Russia\\
% ...
      }
    \end{center}
    \vspace{0.4cm}
  \end{small}
}

\affiliation{}

%% file: draft_Zc0.bbl
\begin{thebibliography}{99}
\bibitem{implied}
Charge conjugation is always implied, unless stated specially.
\end{thebibliography}
